\documentclass[rnaas]{aastex63}

\newcommand{\swift}{\textit{Swift}}
\newcommand{\rf}{\textit{realfast}}

\submitjournal{RNAAS}
\shorttitle{Low-Latency High-Energy Follow-Up}
\shortauthors{Tohuvavohu et al.}
\graphicspath{{./}{figures/}}

\begin{document}

\title{A Demonstration of Extremely Low Latency $\gamma$-ray, X-Ray \& UV Follow-Up of a Millisecond Radio Transient}

\author{Aaron Tohuvavohu}
\affiliation{Department of Astronomy \& Astrophysics, University of Toronto, 50 St. George Street, Toronto, Ontario, M5S 3H4 Canada}

\author{Casey J. Law}
\affiliation{Cahill Center for Astronomy and Astrophysics, MC 249-17 California Institute of Technology, Pasadena, CA 91125, USA}

\author{Jamie A. Kennea}
\affiliation{Department of Astronomy \& Astrophysics, Penn State University, PA, USA}

\author{Elizabeth A. K. Adams}
\affiliation{ASTRON, the Netherlands Institute for Radio Astronomy, Oude Hoogeveensedijk 4, 7991 PD, Dwingeloo, The Netherlands.}
\affiliation{Kapteyn Astronomical Institute, University of Groningen Postbus 800, 9700 AV Groningen, The Netherlands}

\author{Kshitij Aggarwal}
\affiliation{Center for Gravitational Waves and Cosmology, West Virginia University, Chestnut Ridge Research Building, Morgantown, WV 26505}
\affiliation{Department of Physics and Astronomy, West Virginia University, Morgantown, WV 26506}

\author{Geoffrey Bower}
\affiliation{ASIAA, 645 N. A'ohoku Pl, Hilo, HI 96720}
\affiliation{Affiliate Graduate Faculty, UH Manoa Physics and Astronomy}

\author{Sarah Burke-Spolaor}
\affiliation{Center for Gravitational Waves and Cosmology, West Virginia University, Chestnut Ridge Research Building, Morgantown, WV 26505}
\affiliation{Department of Physics and Astronomy, West Virginia University, Morgantown, WV 26506}
\affiliation{CIFAR Azrieli Global Scholars program, CIFAR, Toronto, Canada}

\author{Bryan J. Butler}
\affiliation{National Radio Astronomy Observatory, Socorro, NM, 87801, USA}

\author{John M. Cannon}
\affiliation{Department of Physics \& Astronomy, Macalester College, 1600 Grand Avenue, Saint Paul, MN 55105}

\author{S. Bradley Cenko}
\affiliation{NASA Goddard Space Flight Center, Greenbelt, MD, USA}

\author{James DeLaunay}
\affiliation{Department of Astronomy \& Astrophysics, Penn State University, PA, USA}

\author{Paul Demorest}
\affiliation{National Radio Astronomy Observatory, Socorro, NM, 87801, USA}

\author{Maria R. Drout}
\affiliation{Department of Astronomy \& Astrophysics, University of Toronto, 50 St. George Street, Toronto, Ontario, M5S 3H4 Canada}

\author{Philip A. Evans}
\affiliation{University of Leicester, X-ray and Observational Astronomy Group, School of Physics and Astronomy, University Road, Leicester, LE1
7RH, UK}

\author{Alec S. Hirschauer}
\affiliation{Space Telescope Science Institute}

\author{T. J. W. Lazio}
\affiliation{Jet Propulsion Laboratory, California Institute of Technology, M/S 67-201, 4800 Oak Grove Dr., Pasadena, CA  91109}

\author{Justin Linford}
\affiliation{National Radio Astronomy Observatory, Socorro, NM, 87801, USA}

\author{Francis E. Marshall}
\affiliation{NASA Goddard Space Flight Center, Greenbelt, MD, USA}

\author{Kristen.~B.~W. McQuinn}
\affiliation{Rutgers University, Department of Physics and Astronomy, 136 Frelinghuysen Road, Piscataway, NJ 08854, USA} 

\author{Emily Petroff}
\affiliation{Anton Pannekoek Institute, University of Amsterdam, Postbus 94249, 1090 GE Amsterdam, The Netherlands}

\author{Evan D. Skillman}
\affiliation{Minnesota Institute for Astrophysics, University of Minnesota, 116 Church St. SE, Minneapolis, MN 55455}

\begin{abstract}
We report results of a novel high-energy follow-up observation of a potential Fast Radio Burst. The radio burst was detected by VLA/\rf\ and followed-up by the Neil Gehrels \swift\ Observatory in very low latency utilizing new operational capabilities of \swift\ \citep{guano}, with pointed soft X-ray and UV observations beginning at T0+32 minutes, and hard X-ray/gamma-ray event data saved around T0. These observations are $>10$x faster than previous X-ray/UV follow-up of any radio transient to date. No emission is seen coincident with the FRB candidate at T0, with a 0.2s fluence $5\sigma$ upper limit of $1.35\times10^{-8}$ erg cm$^{-2}$ (14-195 keV) for a SGR 1935+2154-like flare, nor at T0+32 minutes down to $3\sigma$ upper limits of 22.18 AB mag in UVOT \textit{u} band, and $3.33\times10^{-13}$\ erg cm$^{-2}$\ s$^{-1}$\ from 0.3-10 keV for the 2 ks observation. The candidate FRB alone is not significant enough to be considered astrophysical, so this note serves as a technical demonstration. These new \swift\ operational capabilities will allow future FRB detections to be followed up with XRT/UVOT at even lower latencies than demonstrated here: 15-20 minutes should be regularly achievable, and 5-10 minutes occasionally achievable. We encourage FRB detecting facilities to release alerts in low latency to enable this science.
\end{abstract}

\section{Introduction}
\label{sec:intro}

Fast Radio Bursts (FRBs) are a new class of radio transient of unknown origin  \citep{2019A&ARv..27....4P,2019arXiv190605878C}. FRBs have energy releases on the order of $10^{40}\ \rm{erg}$\ in a millisecond and are detectable at cosmological distances \citep{2017Natur.541...58C,2019Sci...365..565B,2020Natur.577..190M,2019Natur.572..352R}. This makes FRBs orders of magnitude more luminous than Galactic pulsars and likely require an entirely different physical mechanism \citep{2014ApJ...797...70K,2016ApJ...826..226K}.

Magnetars are a viable source for FRBs, given the tremendous amount of magnetic energy available through their magnetospheres or shocks driven by relativistic outflows \citep{2019MNRAS.485.4091M,2020MNRAS.494.1217K}. Magnetar models have received new attention with the recent association of an X-ray burst with an FRB-like burst from the Galactic magnetar SGR 1935+2154 \citep{2020arXiv200510828B,2020arXiv200510324T}. FRB models inspired by the SGR 1935+2154 burst predict that FRBs are associated with periods of high-energy activity \citep{2020arXiv200505283M,2020arXiv200506736L}. 
A much stronger magnetar high-energy outburst from SGR 1806--20 did not produce an FRB \citep{2016ApJ...827...59T}. However, outbursts from extragalactic soft-gamma repeaters occur at a significant rate \citep{2005Natur.434.1098H} that is consistent with the observed FRB rate \citep{2020arXiv200510828B}. This argues that FRBs should be directly associated with high-energy bursts and periods of high activity. 
Low-latency high-energy follow-up of FRBs is a viable strategy to test this and other models and characterize the emission process for FRBs.

The NASA Neil Gehrels \swift\ Observatory (\swift; \citealt{2004ApJ...611.1005G}), is a MIDEX class mission with a sensitive and wide field of view (2 sr) hard-X-ray Burst Alert Telescope (BAT; \citealt{BAT}), and has been a discovery engine for new SGRs (e.g. \citealt{2010ApJ...718..331G,Kennea13}). In addition \swift's target-of-opportunity capability allows for rapid repointing of the observatory to observe transients at very low latency (hours to minutes), allowing for accurate localization and study with its narrow field X-ray Telescope (XRT; \citealt{XRT}) and Ultraviolet/Optical Telescope (UVOT; \citealt{UVOT}).

In practice, FRB discovery tends to be limited by telescope sensitivity, so external information can be used to improve sensitivity to astrophysical sources under the assumption of a physical model. The models inspired by SGR 1935+2154 rely on processes with temporally associated radio/x-ray bursts. Furthermore, some models predict persistent high-energy emission associated with FRBs \citep{2018arXiv180809969M}, and there has been a possible association with a long-lived (hundreds of seconds) gamma-ray source in \swift/BAT imaging data \citep{2016ApJ...832L...1D}.

%[Geoff:  It might be worth pointing to some models that don’t expect X-ray association with FRBs.  Also, it would be worth giving a brief summary of X-ray follow-up lags up to this point.] [Aaron:i Dont think this is necessary for this note, honestly. seems too much.]

This note describes the first very low-latency high-energy response to an FRB candidate, utilizing new \swift\ capabilities. The radio event seen by \rf\ had a relatively low significance and no high-energy counterpart was identified. As such, this serves as a technical demonstration of the integration of a low-latency FRB discovery system with \swift.

\section{Candidate FRB}
\rf\ is a system at the Very Large Array (VLA) for commensal fast transient searches \citep{2018ApJS..236....8L}. The VLA produces visibilities at a 10 ms cadence and the \rf\ system calibrates, dedisperses, and images the data to identify fast radio transients in real time. The search typically completes within 3 minutes of the time the radio signal arrives at the telescope. Raw data for candidates are saved for more detailed analysis. The real-time system filters the candidates for FRB-like DM, astrophysical probability and S/N ratio.

On 8 April 2020, the VLA was observing under program 20A-330 (PI: Cannon). \rf\ detected a candidate FRB with a dispersion measure (DM) of 1463.8 pc cm$^{-3}$ and time width 10~ms at MJD 58947.3514950 (topocentric; dedispersed to 2.0 GHz). The candidate was localized to RA, Dec (J2000) $=$ 12h05m24s, +27d53m24.6s (uncertainty of $\sim1$\arcsec; 12\arcsec\ FWHM synthesized beam). The primary VLA was pointed at a nearby, low-mass galaxy (AGC731921; RA, Dec $=$ 12h05m34.3s, +28d13m56s) with a primary goal of studying its HI spectral line and 1-2 GHz continuum properties. The angular offset between the primary target and the FRB candidate is 20.65 arcmin, so they are likely not related.

The commensal transient search included an observing band at frequencies from 1.4 to 2.0 GHz. In this band, the VLA has a nominal sensitivity of 5 mJy in each 10 ms image. The candidate had a S/N ratio of 7.6 in the 10 ms image. This significance implies an approximate fluence of 0.8 Jy ms after primary beam correction (no absolute flux calibration is available). 

The \rf\ system filters candidates for DM, S/N ratio, and astrophysical probability before sending them to \swift. The DM of this event was well in excess of the maximum Milky Way contribution in this direction \citep[20 pc cm$^{-3}$;][]{2002astro.ph..7156C}. The astrophysical score was calculated by fetch \citep{2019arXiv190206343A} to be 0.99, which exceeded the score threshold. Finally, the real-time analysis includes a calculation of whether spectral variations are consistent with noise via a Kalman process estimator (Zackay et al, in prep). The image and spectral variations together are consistent with an 8.0$\sigma$ deviation, which was sufficient to exceed our S/N trigger threshold. An $8\sigma$ signal had a false-alarm rate of approximately 0.1 hr$^{-1}$. This event met all criteria for triggering \swift, so a notice was sent.

\section{High-Energy Follow-Up}
The notice by \rf\ was distributed to \swift\ via a VOEvent listener running on the infrastructure of the Astrophysical Multi-messenger Observatory Network \citep{amon}, using the standard for Fast Radio Bursts \citep{petroffvoevent}. Receipt of this notice at the \swift\ Mission Operations Center (MOC) triggered the GUANO system \citep{guano} to autonomously recover the BAT event data around T0, as well as an automatic, highest urgency, Target of Opportunity (ToO) request to point the XRT and UVOT.\\

\noindent
T0 =  2020-04-08 08:26:07.65 UTC: Radio transient time de-dispersed to infinite frequency.\\
T0+5 minutes: Candidate identified by \rf.\\
T0+6 minutes: Notice received at \swift\ MOC.\\
T0+29 minutes: GUANO command received onboard \swift, BAT ring buffer dumped.\\
T0+29.3 minutes: ToO repoint command received onboard \swift. \\
T0+29.5 minutes: Slew towards FRB location begins (100.4 degrees to target).\\
T0+32 minutes: \swift\ settled on target, XRT and UVOT observations begin.\\

% From Casey: dispersion correction between top of radio band (2ghz) to infinite frequency -->   4.1488e-3*1463.8*(1/2**2) = 1.518s

\begin{figure}[ht!]

 $\vcenter{\hbox{\includegraphics[width=0.5\textwidth]{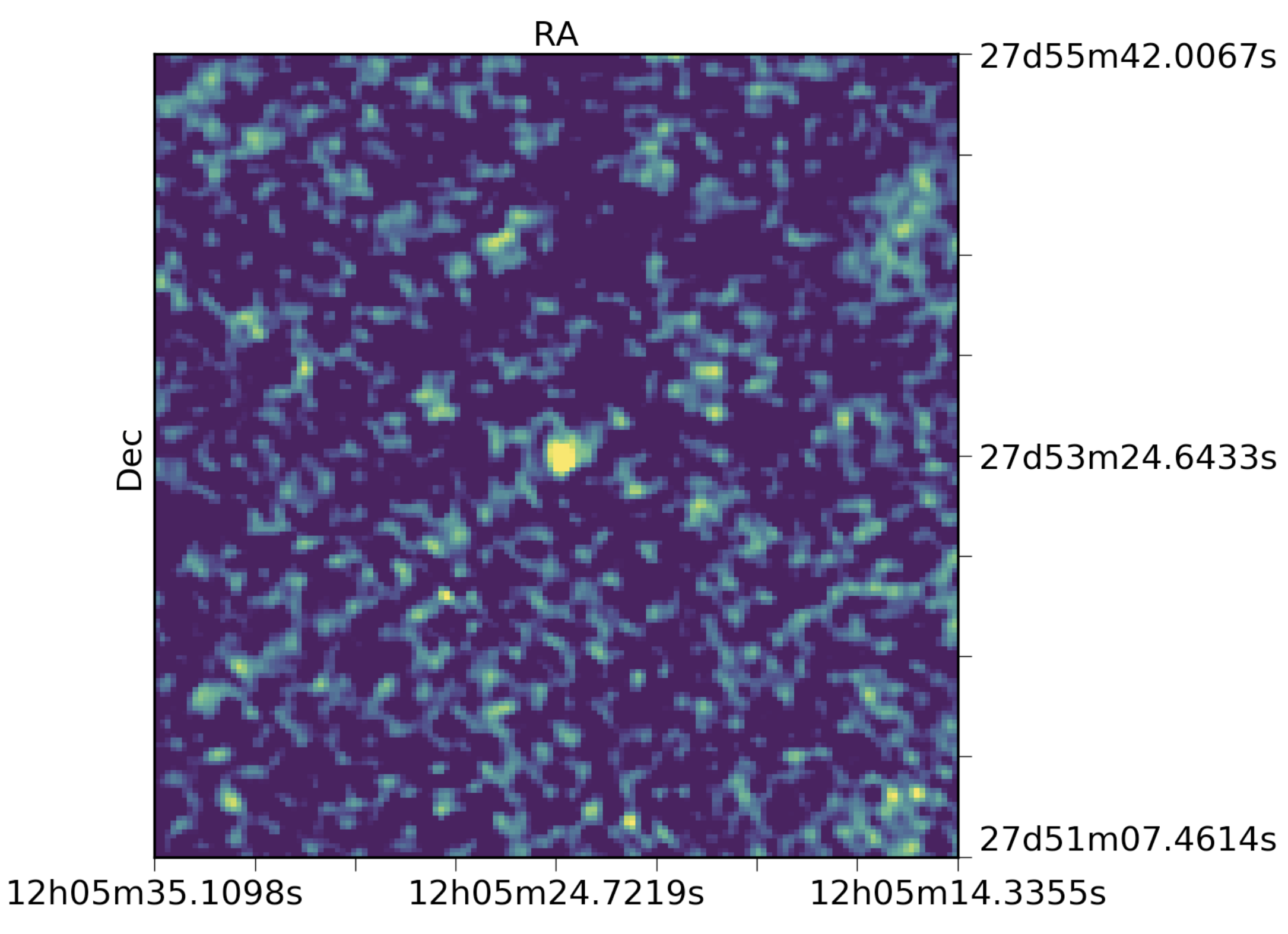}}}$
  \hspace*{.2in}
  $\vcenter{\hbox{\includegraphics[width=0.4\textwidth]{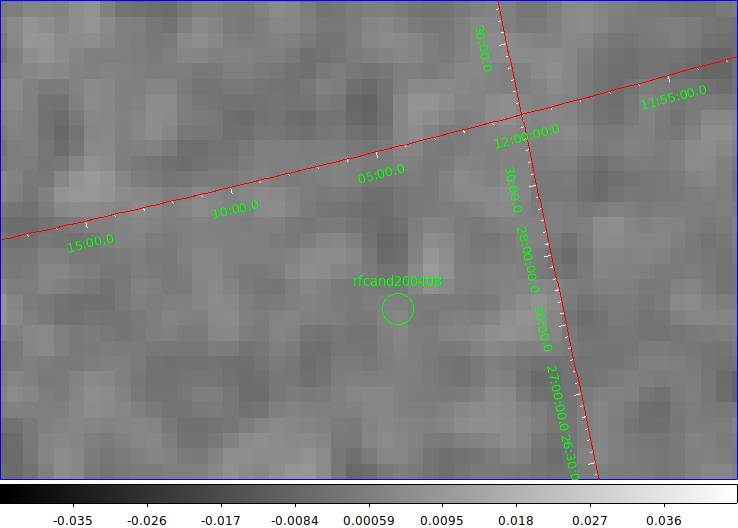}}}$
    
\caption{Left: VLA radio image of FRB candidate. Right: BAT sky image (15-300 keV) of the location of the radio transient at dedispersed T0. \label{fig:data}}
\end{figure}
%utcf at T0 was 24.7394 seconds
%Made a BAT sky image centered at bat MET time of 608027192.3914274 (notice trigger time + utcf - dispersion correction)

At T0, \swift\ was slewing between pre-planned targets, but the location of the radio transient was in the contemporaneous field-of-view of the BAT, at a partial coding fraction of $75\%$. Using the event data made available via GUANO, we searched for a prompt gamma-ray transient counterpart to the radio transient. The analysis is complicated beyond the normal BAT analysis by the fact that \swift\ was performing a slew maneuver at T0. We constructed Detector Plane Images, and then sky images, for every 0.2 second time bin during the slew (the temporal resolution at which attitude information is available). We searched for sources in the sky images with \texttt{batcelldetect}, and found no source at the position of the radio transient. We extracted the upper limit in BAT counts for the 0.2s duration noise map centered around the de-dispersed radio transient T0, and converted it to a fluence upper limit assuming a power law spectrum with a photon index of $\Gamma=2.1$, as found for the X-ray flare associated with FRB 200428 \citep{mereghetti2020integral}. We find a $5\sigma$ fluence upper limit of $1.35\times10^{-8}$ erg cm$^{-2}$ (14-195 keV).  Assuming instead a power law with a photon index of $\Gamma=-1.32$ (typical for short GRBs in the BAT band; \citealt{Lien_2016}) the $5\sigma$ fluence upper limit becomes $3\times10^{-8}$ erg cm$^{-2}$ (14-195 keV) for the 0.2s duration around T0.

UVOT took an exposure of 2 ks seconds in event mode (11 ms temporal resolution) with the \textit{u} filter ($\lambda_{center}=$ 346 nm) beginning at T0+32 minutes. We screened and cleaned the event list, before using the \swift\ ftools task \texttt{uvotimage} to stack into a sky image. The UVOT image was then analyzed using the standard \texttt{uvotsource} task, which performs aperture photometry using user-specified source and background regions. A 5" aperture was used for the source region and a nearby source-free region was chosen for the background. Using the UVOT photometric system \citep{Breeveld2011} we compute a 3-sigma upper limit of $u> 22.18$ mag AB, for any source at the location of the FRB candidate.

XRT accumulated 2 ks of cleaned exposure time in Photon Counting mode beginning at T0+32 minutes. Using the tools of \citep{philonline} we compute a 3-sigma upper limit of $6.0\times10^{-3}$ ct s$^{-1}$. Assuming a spectrum that can be modelled with an absorbed power law with $\Gamma$ of 1.2 (as found for the post-flare soft X-ray emission of SGR 1935+2154 in \citealt{borghese2020xray}) and a galactic $N_H$ \footnote{Clalculated with the \texttt{nhtot} tool \citep{Willingale2013}.} of $1.76\times10^{20}$ cm$^{-2}$, this corresponds to a 3-sigma flux upper limit of $3.33\times10^{-13}$ erg cm$^{-2}$ s$^{-1}$ from 0.3-10 keV.

\section{Conclusions}
This note describes the fastest X-ray/UV follow-up of any radio transient to date. The radio signal had a relatively low significance and no independent high-energy counterpart was identified; the event was unlikely to be astrophysical.

While in this case XRT/UVOT began observations at T0+32 minutes, even lower latency observations are achievable in the future. The GUANO system and the autonomous commanding developed for it has now achieved average commanding latencies of 14 minutes with even lower latencies possible some of the time (see latency distribution in Figure 6 of \citealt{guano}). 
% Added by JAK 6/2/20
The combination of GUANO commanding and fully automated ToO safety checking and scheduling being developed for possible use by \swift\ should allow observations of FRB candidates with the focused narrow field instruments to begin within typically 15-17 minutes (i.e. 14 minutes plus the time to slew to target) in many cases, depending on visibility constraints at T0. If a FRB triggers near in time to an already existing \swift\ commanding pass, observations could begin in as little as $\sim5$ minutes. The current follow-up program is limited only by the triggering rate of \rf. We estimate that at a trigger rate of $\sim$ a few FRBs per day (achievable by current instruments like CHIME/FRB \citealt{CHIMEFRB}), XRT/UVOT observations could begin with a latency of $<\sim10$ minutes for $\sim$ one trigger per month. We encourage FRB facilities to report their triggers in low-latency to enable this science, which are necessary to probe progenitor models with rapidly fading high-energy afterglows.

\vspace{5mm}
\facilities{Karl G. Jansky Very Large Array, Neil Gehrels \swift\ Observatory}

\software{rfpipe \citep{2017ascl.soft10002L}, comet \citep{2014A&C.....7...12S}, fetch \citep{2019arXiv190206343A}, GUANO \citep{guano}}

\acknowledgements
This work was supported by NASA \swift\ Guest Investigator grant 80NSSC20K0945, and by the National Science Foundation under grants PHY-1708146 and PHY-1806854 and by the Institute for Gravitation and the Cosmos of the Pennsylvania State University. This work made use of data supplied by the UK Swift Science Data Centre at the University of Leicester. CJL acknowledges support under NSF grant 2022546. Part of this research was carried out at the Jet Propulsion Laboratory, California Institute of Technology, under a contract with the National Aeronautics and Space Administration.
The NANOGrav project receives support from National Science Foundation (NSF) Physics Frontiers Center award number 1430284.

\bibliography{main}{}
\bibliographystyle{aasjournal}

\end{document}